\documentclass[aps,prb,reprint,amssymb,amsfonts]{revtex4-1} 
\usepackage{mathptmx} 
\usepackage{bm}
\usepackage{graphicx}
\usepackage{color}
\usepackage{hyperref}
\hypersetup{
pdfborder={0 0 0}, 
colorlinks=true, 
linkcolor=blue,
citecolor=blue,
urlcolor=blue
}

\begin{document}
\title{Identification of Kelvin waves: numerical challenges}
\author{R. H\"anninen}
\affiliation{O.V. Lounasmaa Laboratory, Aalto University, PO BOX 15100, 00076 AALTO, Finland}
\author{N. Hietala}
\affiliation{O.V. Lounasmaa Laboratory, Aalto University, PO BOX 15100, 00076 AALTO, Finland}

\date{December 11, 2012}

\begin{abstract}
Kelvin waves are expected to play an essential role in the energy dissipation for quantized vortices. However, the 
identification of these helical distortions is not straightforward, especially in case of vortex tangle. Here we 
review several numerical methods that have been used to identify Kelvin waves within the vortex filament model. 
We test their validity using several examples and estimate whether these methods are accurate enough to verify 
the correct Kelvin spectrum. We also illustrate how the correlation dimension is related to different Kelvin spectra
and remind that the 3D energy spectrum $E(k)$ takes the form $1/k$ in the high-$k$ region, even in the presence of 
Kelvin waves. 
\end{abstract}

%
\maketitle


\section{Introduction}

Dissipation of energy in the zero temperature limit is a central question in the field of quantum turbulence.
In helium superfluids turbulence is related to motion of quantized vortices that generate the superfluid velocity
field. At scales smaller than the intervortex distance energy is expected to be cascaded to smaller scales 
via Kelvin waves (KWs) until it can be transformed into thermal excitations, phonons in case of $^4$He.
\cite{SvistunovPRB1995,VinenJLTP2002}
Therefore, the proper identification of Kelvin waves is important if we want to verify the theoretically 
predicted Kelvin spectrum\cite{KS2004prl,LN2010,Sonin2012}. Experimentally this is probably not possible, 
but numerically the identification is conceivable. Actually, several numerical simulations have tried to verify 
the correct spectrum using the vortex filament model\cite{VinenPRL2003,KS2005prl,KS2010prbsub,BaggaleyPRB2011}.
In our opinion, none of those is fully convincing. The cascade towards smaller scales is evident in simulations 
but the correct value for the slope of the spectrum is dominated by a numerical noise or distorted by the method 
used to determine the Kelvin amplitude. The simulations by Kozik {\it et al.}\cite{KS2005prl,KS2010prbsub}, 
using the Hamiltonian formulation of the filament model, correctly identify the Kelvin modes, but their spectrum 
may depend on the absolute amplitude of the initial spectrum\cite{Sonin2012,HanninenPin}. 

In BEC's the Kelvin spectrum has been determined by Yepez {\it et al.}\cite{YepezPRL2009} using the Gross-Pitaevskii 
equation. However, the interpretation of these results was strongly criticized\cite{Commentklow,Commentk3}. 
In high-$k$ region the $k^{-3}$ spectrum (originally interpeted as Kelvin spectrum) most likely results from 
the velocity and density profile inside the vortex cores\cite{Commentk3}. The low-$k$ region was originally 
interpreted as Kolmogorov K41 law, even if the spectrum is likely visible at scales smaller than the intervortex 
distance (authors did not evaluate the mean vortex separation). One option is that this spectrum comes from Kelvin 
waves, but since the fluid is highly compressible the explanations presented in the Comment\cite{Commentklow} 
are more likely.

In this article we list several ways that can be used to identify Kelvin waves. We work in the framework of the 
vortex filament model\cite{schwarz85}, where the vortices are considered to be thin and the superfluid 
velocity can be calculated simply from the vortex configuration ${\bf s}(\xi)$, where $\xi$ is the length 
along the vortex, using the Biot-Savart integral. In other words, we consider only length scales much larger than 
the vortex core size, $a$. We start with two simple cases where the definition of the Kelvin waves is obvious:
straight vortex with Kelvin waves and a vortex ring occupied by KWs. In both cases the Kelvin spectrum can be 
determined using a simple Fourier transformation. We use these two sample cases to test other methods that
have been used previously in the literature. Such methods are, for example, curvature, energy spectrum and
fractal dimension.

\section{Kelvin waves}

Kelvin waves are helical distortions on a vortex. 
They can be generated by applying a counterflow ${\bf v}_n-{\bf v}_s$ along a vortex\cite{GlabersonPRL1974}.
Vortex reconnection, and more generally, interaction 
with other vortices, with vortex itself or with boundaries also induce Kelvin waves. In the small amplitude 
and long wavelength limit the dispersion relation for a Kelvin wave with wave vector $k$ is given by
\begin{equation}\label{e.kelvinfreq}
\omega_k = \frac{\kappa k^2}{4\pi}\lbrack \ln(2/ka) - \gamma \rbrack\,.
\end{equation}
Here $\kappa = h/m_4$ (for $^4$He) or $h/2m_3$ (for $^3$He-B) is the circulation quantum, $a$ the vortex 
core size and $\gamma = 0.57721\ldots$ is the Euler constant. 

On a straight vortex (taken to be along the $z$-axis) the definition and identification of the Kelvin waves is 
simple. Assume that a vortex can be represented as coordinates $x(z)$ and $y(z)$ and define 
$w(z) = x(z)+iy(z)$. Now the vortex configuration can be presented as a sum of different Kelvin modes:
\begin{equation}\label{e.kwspect}
w(z) = \sum_k w_k\exp({\rm i} k z)\,.
\end{equation}
In case of periodic boundary conditions along the $z$-direction (typical for simulations), the $k$-vector 
is discrete and given by $k = 2\pi{m}/L_z$, where $m$ (mode) is integer and $L_z$ is the period.  
For a particular vortex configuration Kelvin spectrum can be defined as $n_k=|w_k|^2+|w_{-k}|^2$ with $k>0$.
Theoretical predictions, which apply for a statistical average, predict that the steady-state 
Kelvin spectrum takes the form\cite{KS2004prl,LN2010,Sonin2012}
\begin{equation}\label{e.kspec}
n_k=|w_k|^2+|w_{-k}|^2 = N_0 k^{-2\eta}\,,\,\hspace{5mm} (k>0)
\end{equation}
with $\eta \approx 1.5\ldots 2$. This spectrum results from interaction of different scale Kelvin 
waves and provides a constant ($k$-independent) energy flux, $\theta$, through different scales.
If $\epsilon_m$ (with $m>0$) determines the fraction of positive
Kelvin modes, then we can parametrize the amplitudes using the modes:
\begin{eqnarray}
|w_m| &=& \sqrt{\epsilon_m}\,A\,m^{-\eta}, \\
|w_{-m}| &=& \sqrt{1-\epsilon_m}\,A\,m^{-\eta}. \nonumber 
\end{eqnarray}

Numerically, a great care should be used when identifying the spectrum. If the vortex configuration
is such that the local curvature approaches the numerical resolution, $k_{\rm res}$, then even the interpolation 
used to obtain the vortex configuration at equidistant points (needed for FFT) can result in errors that 
distort the spectrum for $k$-values $k \gtrsim k_{\rm res}/5$. One way to avoid these errors
is to directly solve the Fourier coefficients. But then one loses the speedup obtained by FFT. 
An alternative is to keep the vortex points equally $z$-distributed by removing the motion along 
the $z$-direction. By introducing a vector ${\bf \rho}(z,t) = (x(z,t),y(z,t))$, which only determines
the point of intersection of the vortex line with the plane, one obtains that\cite{SvistunovPRB1995,KozikJLTP2009}
\begin{equation}\label{e.rho}
\frac{\partial{\bf \rho}}{\partial t} = \frac{\partial{\bf s}}{\partial t}-
\left( \hat{\bf z}\cdot \frac{\partial{\bf s}}{\partial t}\right )
\left( \hat{\bf z}+\frac{\partial{\bf \rho}}{\partial z} \right)\,.
\end{equation}
Here $\hat{\bf z}$ is the unit vector along the $z$-axis. This formula, together with the Biot-Savart
equation can be used to derive the Hamiltonian equation for the vortex motion\cite{SvistunovPRB1995,KozikJLTP2009}. 
This projection should be used with care, since it limits the real vortex motion. It, for example, prohibits 
a vortex to reconnect with itself.  

Cascade due to Kelvin waves is typically very weak. The energy flux, $\theta$, can be increased by 
increasing the amplitudes of the Kelvin waves. 
Simulations always contain some dissipation (or noise) that typically is strongest at the smallest scales. 
For small energy flux this numerical dissipation may, in some cases, correctly mimic the energy sink. However, 
if the numerics is done in a such way that the energy is well conserved, this flow out from the numerical $k$-region can 
become smaller than the energy flux due to Kelvin-wave cascade. This results in a numerical bottleneck, 
implying accumulation of Kelvin waves near the resolution limit and generally resulting excessive fractality 
and noise. In simulations this can often be seen as a failure of the Fourier presentation for the vortex. Therefore, 
any numerical method that tries to determine the steady-state Kelvin spectrum should contain enough dissipation 
such that this bottleneck due to finite resolution is avoided. How this is implemented in numerics, especially 
with the vortex filament model, is still somewhat open question. One option is to work at finite temperatures 
and have high enough resolution such that all the energy can be dissipated by mutual friction. Important is, 
that a proper dissipation is arranged dynamically. Simply damping the highest modes might not be enough.

One should also note that near the resolution limit the dispersion relation, Eq.~(\ref{e.kelvinfreq}),
may become distorted. Simple estimation states that $\omega$ becomes flat near $k_{\rm res}$. Because
the dynamics is not accurately modeled at the smallest scales, these scales can result in different
spectrum and should therefore be omitted from the fits.

\section{Kelvin waves on a vortex ring}

For a vortex ring with Kelvin waves one can also make a Fourier presentation. Consider a vortex ring 
of radius $R_0$ in the $z=z_0$ plane located symmetrically around the $z$-axis. If we occupy 
the ring with Kelvin waves, in a similar fashion than a straight vortex, we may parametrize it as 
follows:
\begin{eqnarray}\label{e.ringKWs}
x &=& \sum_m R_m\cos(m\phi+\varphi_m)\cos\phi \nonumber \\
y &=& \sum_m R_m\cos(m\phi+\varphi_m)\sin\phi \\
z &=& z_0 -\sum_m R_m\sin(m\phi+\varphi_m) ,\nonumber
\end{eqnarray}
where $\phi = \arctan(y/x)$ is the azimuthal angle and $\varphi_m$ is the phase of the mode $m$. 
Amplitudes $R_m$ (which correspond to $|w_m|$, in case of a straight vortex) can be calculated 
using a FFT in the following way: First the points along the vortex must be interpolated at 
equidistant $\phi$ points. This can be realized using {\it e.g.} cubic interpolation. Now the 
Fourier transformation of
\begin{eqnarray}
Z = x\cos\phi+y\sin\phi-{\rm i}z
\end{eqnarray}
directly determines the $R_m$'s. The zero frequency term from the Fourier transformation gives both 
the ring radius $R_0$ (real part) and also the location of the ring, $z_0$ (imaginary part). 

If we want to occupy the vortex with a particular spectrum, we can take 
\begin{eqnarray}
R_m &=& \sqrt{\epsilon_m}\,A\,m^{-\eta}, \\
R_{-m} &=& \sqrt{1-\epsilon_m}\,A\,m^{-\eta}. \nonumber 
\end{eqnarray}
Here $\epsilon_m$ (with $m>0$) determines the fraction of positive and negative Kelvin modes. 

At zero temperature, or when the normal fluid is at rest, kinetic energy and momentum are 
only due to superfluid component and completely determined by the vortex configuration (we ignore
externally applied velocities). Additionally, if vortices form loops and the vorticity disappears 
at the infinity, one may evaluate kinetic energy, $E$, and momentum, $P$, in numerically convenient 
way by line integrals over the vortex configuration:
\begin{eqnarray}\label{e.enemom}
E &=& \rho_s\kappa\oint {\bf v}_s\cdot {\bf s}\times d{\bf s} \\
P &=& \frac{1}{2}\rho_s\kappa\oint {\bf s}\times d{\bf s} 
\end{eqnarray}
These formulas make it possible to evaluate the accuracy of the numerical scheme used. At zero
temperature both the energy and momentum should be conserved. However, even if the energy is
well conserved, {\it e.g.} by accuracy of 0.1 percent, this fluctuation is still very large if we 
compare it with the energy related to smallest scale Kelvin waves. Therefore, the errors coming from 
the time iteration scheme can be large at the smallest length scales. In the following we are not
concentrating on these errors but pay attention only to methods that try to identify the 
Kelvin waves on a tangle. In other words, we assume that the vortex configuration is correct
at the numerical points ${\bf s}_i$, $i=1,\ldots,N$.

\section{Curvature}

\begin{figure}[!t]
\includegraphics[width=0.8\linewidth]{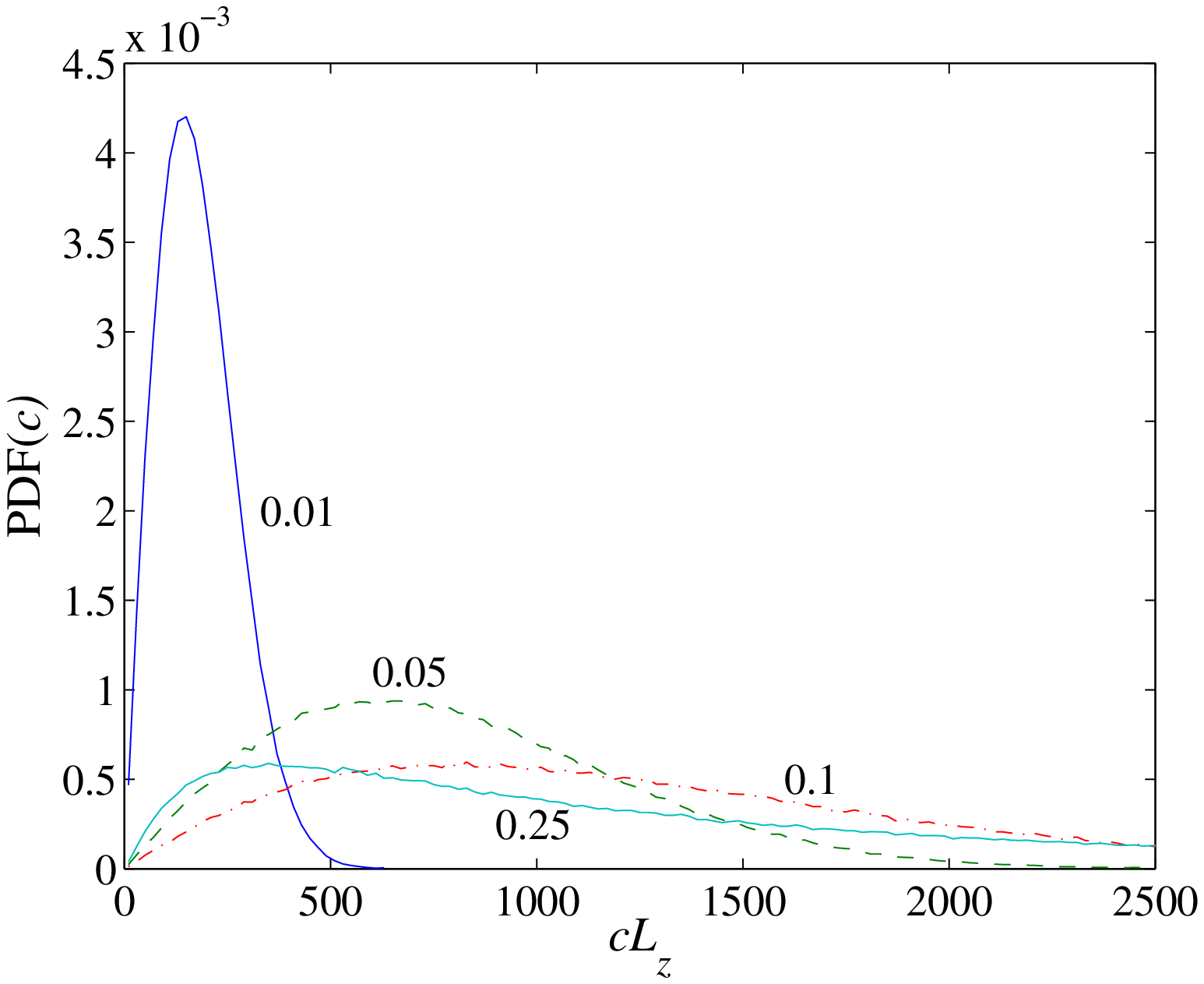}
\includegraphics[width=0.8\linewidth]{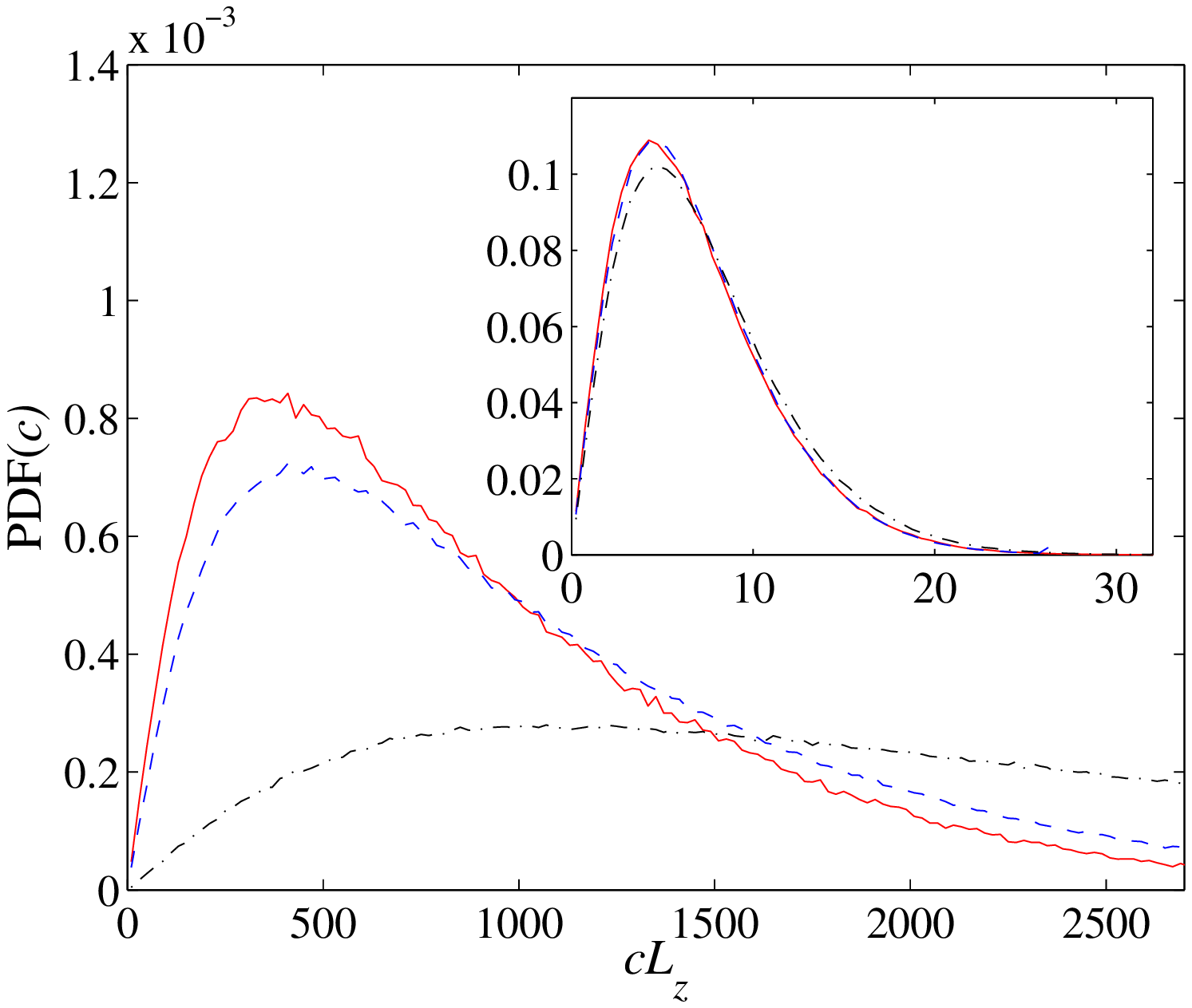}
\caption{(Color Online) Curvature histogram obtained for a straight vortex with Kelvin waves. Histograms 
are averages over 50 different sets, where the phases and the fraction of positive and negative modes 
is randomly set. 
{\it Top:} Curvature histogram for a Kelvin spectrum with $\eta = 1.7$ and having amplitudes $A/L_z$ = 0.01, 
0.05, 0.1, 0.25. Number of points is $N$ = 8192 and all the modes with $|m|>5$ are present.
{\it Bottom:} Curvature histogram for $A = 0.1L_z$. The solid (red) line is calculated with $N$ = 8192 points 
and using all modes (from $|m|=1$ up to the resolution). The dash-dotted (black) line is calculated with four times better 
resolution ($N$ = 32768), also using all modes. The dashed (blue) line is also with $N$ = 32768, but having 
only the same modes than in the low resolution run. The main figure is for $\eta$ = 1.7 and the inset 
is for $\eta$ = 2.5. 
}
\label{f.curvhist}
\end{figure}

In case of vortex tangle the identification of Kelvin waves becomes tedious. One easily seeks
different indirect methods to identify small scale Kelvin waves. The most obvious thing is to 
calculate local curvature $c(\xi)=|{\bf s}''|$ along the vortex. An increasing average curvature is a 
clear sign that more small scale structures appear. A histogram describing the distribution of 
curvature gives some more information, but our tests indicate that curvature histogram is a poor 
method to determine the Kelvin spectrum (exponent $\eta$). For example, the location 
of the histogram maximum depends, not only on the spectrum, but also on the amplitude, as illustrated 
in Fig.~\ref{f.curvhist}. The location of this maximum, as a function of amplitude, is also 
non-monotonous. Similar effect is also seen if one changes $\eta$.

At high curvature regions the histogram has been previously observed (with somewhat limited 
resolution) to take a form $c^{-\alpha}$.\cite{BaggaleyPRB2011} At least our test with spectrum given
by Eq.~(\ref{e.kwspect}), indicate that this kind of fit is poor (exponential fit is often better) 
and the obtained $\alpha$ is very sensitive to the fit region. More importantly, the location
of the maximum and also the shape of the histogram at high curvature regions depends on the 
resolution for spectra $\eta \lesssim 2.5$, as seen in Fig.~\ref{f.curvhist}.  

In case of a straight vortex with small amplitude Kelvin waves one can relate the curvature
spectrum with the Kelvin spectrum in the following way: \cite{KS2004prl} 
\begin{equation}\label{e.curvspect}
\int|c(\xi)|^2d\xi \simeq \int\langle |w''(z)|^2 \rangle dz = \sum_k k^4|w_k|^2\, .
\end{equation} 
Therefore one might expect that the curvature spectrum could reveal the Kelvin spectrum, also
in more complicated tangles. We observed that the curvature spectrum follows this law only
when the amplitude is small and $\eta > 2$, {\it i.e.} only when the curvature spectrum $\propto k^y$ 
is such that $y<0$. For any smaller $\eta$ the curvature spectrum becomes flat and nothing can be 
said about $\eta$. Improving the numerical scheme for determining the local curvature does not resolve 
this problem. If the Kelvin amplitude is not small, the above relation, Eq.~(\ref{e.curvspect}), is 
satisfied even less accurately.

\section{Energy spectrum}

Energy (or velocity) spectrum $E(k)$ describes the distribution of the kinetic energy
at various length scales. At absolute zero temperature the total kinetic 
energy is given by superfluid component 
$E_{k}=E_{k,s}=\frac{1}{2}\rho_s\int v_s^2d^3r = \int E(k)dk$. Where now ${\bf k}$ is the three-dimensional 
vector in the momentum space and $k=|{\bf k}|$. 
(Note that in case of Kelvin waves the $k$-vector was 1D object.)
Assuming a quantized vortex with singular distribution of vorticity and that the vorticity disappears
at the infinity one may derive following formula for the kinetic energy caused by the vortices\cite{ArakiPRL2002}:
\begin{eqnarray}
E_{k,s}= \frac{\rho_s\kappa^2}{2(2\pi)^3}\int\frac{d^3{\bf k}}{k^2}\int d\xi_1 
\int d\xi_2 {\rm e}^{{\rm i}{\bf k}\cdot({\bf s}_1-{\bf s}_2)}
(\hat{\bf s}_1'\cdot\hat{\bf s}_2'). 
\label{e.ekins}
\end{eqnarray}
Here the double integration is along vortices, described by coordinates $\xi_1$, and $\xi_2$. 
Notations ${\bf s}_i = {\bf s}(\xi_i)$ and $\hat{\bf s}_i' = d{\bf s}(\xi_i)/d\xi_i$ are the 
location of the vortex core and tangent at $\xi_i$, respectively. 
Additionally, one should emphasize that this formulation is exact only 
if the vortices form closed loops and that the vorticity disappears far 
away. Otherwise, at least the low $k$-values are calculated incorrectly, 
which is not emphasized in Refs.~[\onlinecite{ArakiPRL2002,NemirJLTP2002,ArakiJLTP2002}].

Fortunately, the integration over different $k$-directions in Eq.\ (\ref{e.ekins}) 
can be done analytically\cite{KondaurovaJLTP2005}, resulting that
\begin{eqnarray}
E(k) = \frac{\rho_s\kappa^2}{(2\pi)^2}\int d\xi_1 
\int d\xi_2 \frac{\sin(k\vert {\bf s}_1-{\bf s}_2\vert)}
{k\vert {\bf s}_1-{\bf s}_2\vert}
(\hat{\bf s}_1'\cdot\hat{\bf s}_2') .
\label{e.ek}
\end{eqnarray}
For anisotropic situations, the dependence on the absolute value of the wave number, $k$, 
should be understood as an angle average.

In our numerical scheme vortices are described by sequence of points and we assume that between 
these points vortex is straight. This is a standard assumption used also by several other 
authors. The numerical integration of the above double integral is quite straightforward 
but, due to the oscillating part of the integrand, some care should be taken when $k$ is large. 
However, we found that it is important to check this limit, because the energy spectrum at 
$k$-values that approach and exceed the numerical resolution limit should converge towards 
$k^{-1}$ spectrum. This is not the case in some of the previous vortex filament calculations 
presented so far \cite{ArakiPRL2002,KondaurovaJLTP2005,KivotidesPRL2001} and therefore some 
caution should be kept in mind when reading those papers. Additionally, one numerical check for 
the spectrum is given by the absolute amplitude of the spectrum: in the high-$k$ region, 
near the resolution, the amplitude should be given by the vortex length. 

\begin{figure}[b]
\includegraphics[width=0.75\linewidth]{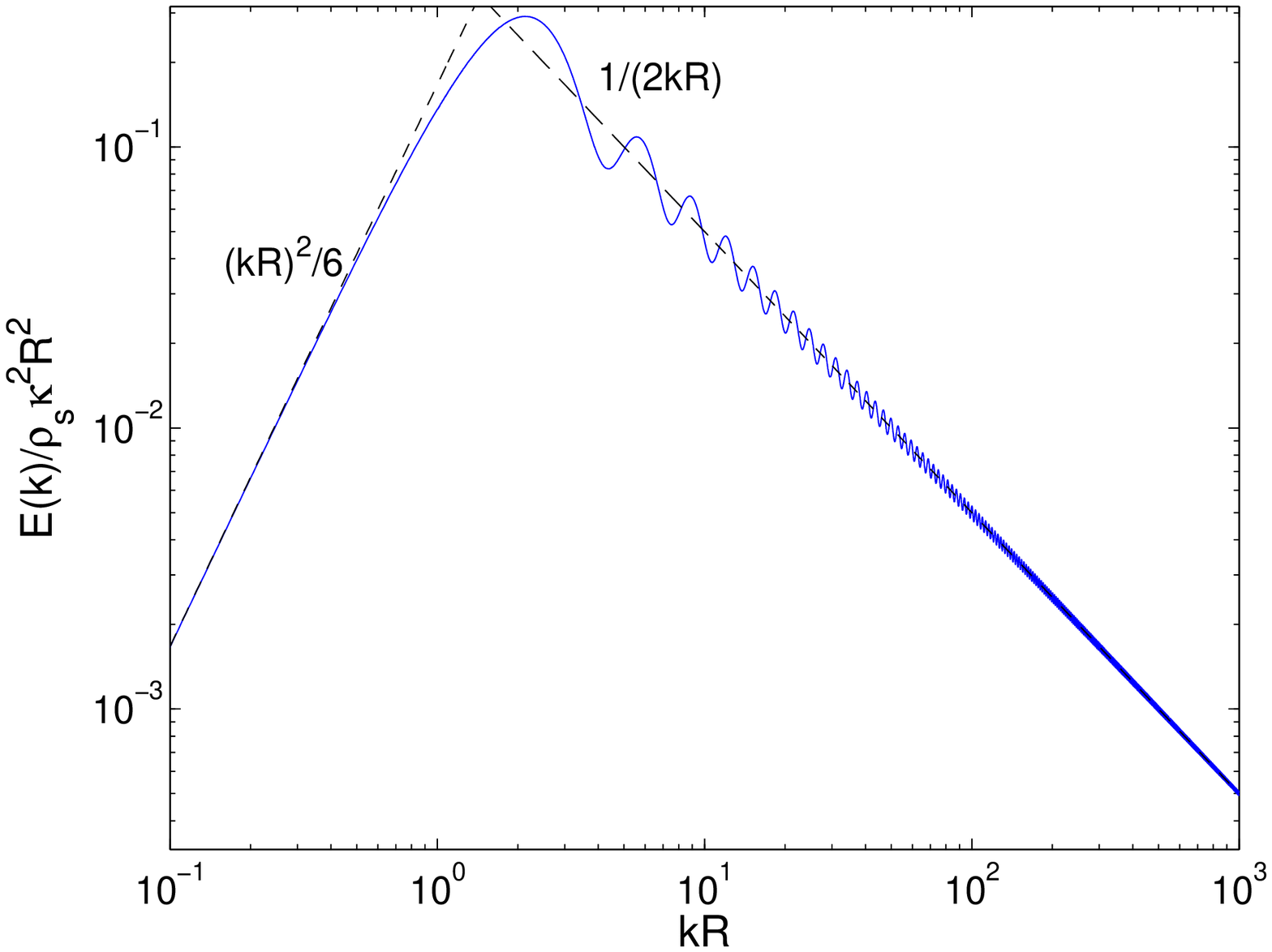}
\includegraphics[width=0.75\linewidth]{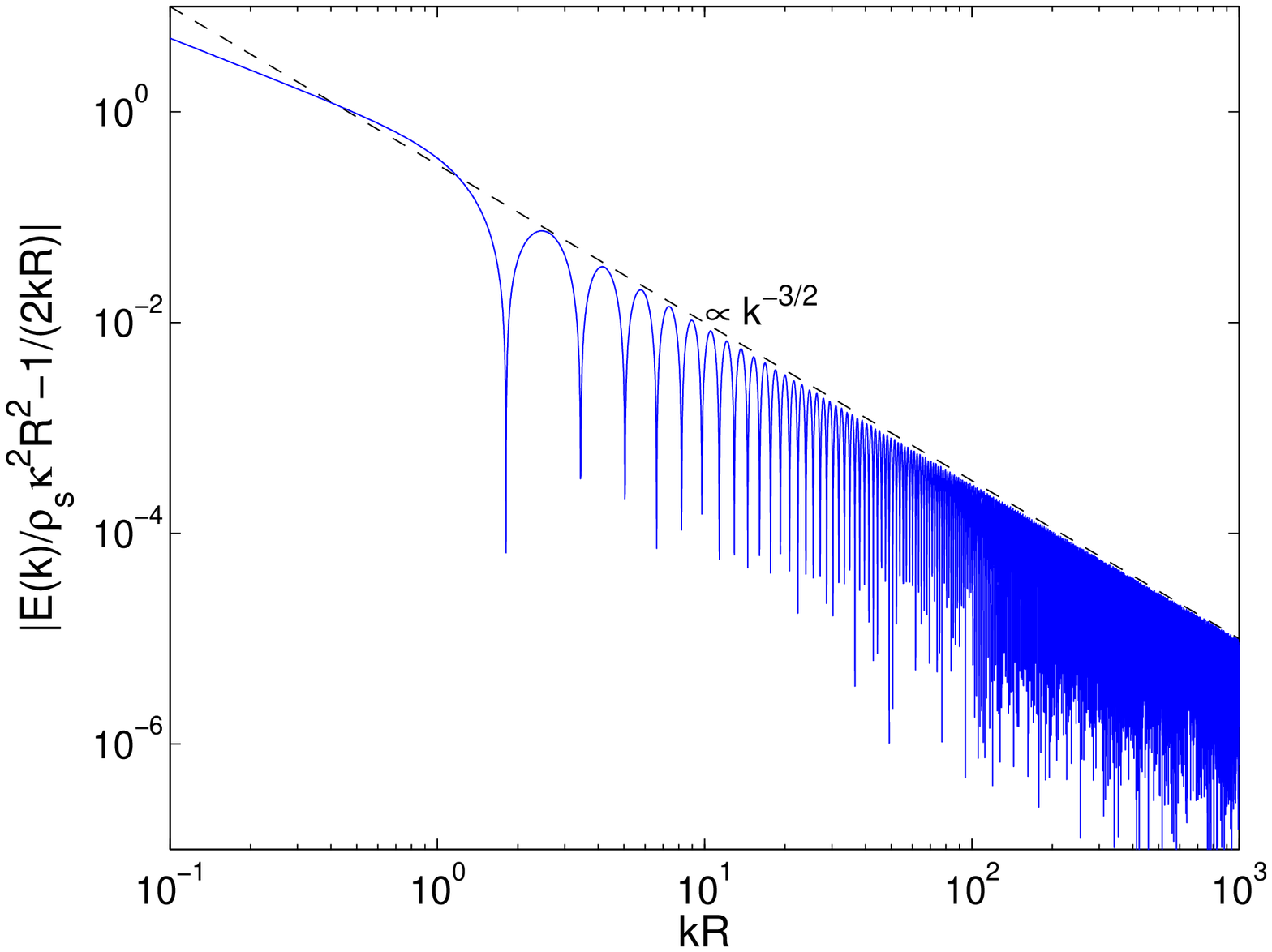}
\caption{(Color Online) {\it Top:} Energy spectrum $E(k)$ of vortex ring of radius $R$. 
At large scales ($kR\ll 1$) $E(k)=\rho_s\kappa^2R^4k^2/6$ and 
at small scales ($kR\gg 1$) $E(k)=\rho_s\kappa^2 R/(2k)$, indicated by the dashed lines.
{\it Bottom:} Deviation of the energy spectrum from the $kR \gg 1$ limit.
The amplitude of the oscillations scales like $k^{-3/2}$.
}
\label{f.ekring}
\end{figure}

For a single vortex ring with radius $R$, the line integration 
formula, Eq.\ (\ref{e.ek}), can be simplified further. Simply putting 
${\bf s}_i=R(\cos\phi_i \hat{\bf x}+\sin\phi_i \hat{\bf y})$ when 
$\hat{\bf s}_1'\cdot\hat{\bf s}_2'=\cos(\phi_1-\phi_2)$ one obtains,
after few standard integration techniques, that
\begin{equation}\label{e.ekring}
E(k) = \frac{\rho_s\kappa^2R^2}{\pi}\int_0^\pi dx\frac{\sin(2kR\sin{x})}{2kR\sin{x}}\cos(2x).
\end{equation}

This can be presented using generalized hypergeometric functions, but we found that 
it is much more convenient to do the integration numerically. The spectrum for vortex 
ring is presented in Fig.\ \ref{f.ekring}. At large scales the energy spectrum is 
given by $E(k)=\rho_s\kappa^2R^4k^2/6$ which can be obtained by doing a series expansion 
at small $k$ for Eq.\ (\ref{e.ekring}). At small scales the vortex ring looks like a 
straight vortex (of length $2\pi R$) and the energy spectrum is given by 
$E(k)=\rho_s\kappa^2 R/(2k)$. The spectrum for single ring is very similar to the one 
produced by four rings in Ref.~[\onlinecite{KivotidesPRL2001}], but our better resolution makes it 
possible to observe small amplitude oscillations. We found that these oscillations,
with period $\pi/R$, around this limiting curve scale like $k^{-3/2}$, which is illustrated 
in Fig.\ \ref{f.ekring}. 
To resolve these oscillations the discretization step for 
different $k$-values should be smaller than this oscillation period, which makes 
the calculations of high $k$-values time consuming. 

An advantage of the line integration formula over the standard Fourier transformation 
used {\it e.g.} by Kivotides {\it et al.}\cite{KivotidesPRL2001} is that one avoids the numerical 
problems that is caused by the divergent $1/r$ velocity field generated by vortices. 
A disadvantage of this method is that in the presence of the walls or with periodic 
boundaries some care should be taken since not all of the assumptions in deriving 
Eq.~(\ref{e.ekins}) are fulfilled. Typically the boundaries result in extra surface 
terms that make the correct calculation more cumbersome. For a single plane boundary 
the energy spectrum can be calculated by extending the second line integral
in Eq.~(\ref{e.ek}) to include also the image vortices. Nevertheless, the high 
$k$-values are correctly taken into account even with this simple formulation.

\begin{figure}[!t]
\centerline{\includegraphics[width=0.8\linewidth]{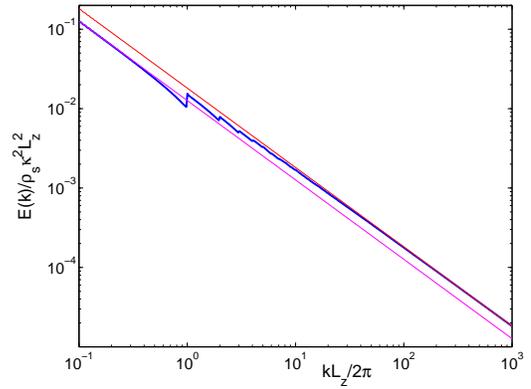}}
\caption{(Color Online) Energy spectrum $E(k)$ of a straight vortex with Kelvin waves. The spectrum
for Kelvin waves is given by $\eta$ = 1.7 with amplitude $A = 0.1L_z$, where $L_z$ = 1 mm
is the period along $z$-direction. The two straight lines corresponds spectrum of straight 
vortex with length $L=L_z$ (lower) and $L=1.4347L_z$ (upper), which is the total length 
of the vortex (per period used).}  
\label{f.EkVortexKWs}
\end{figure}

Even if the 3D energy spectrum $E(k)$ is suitable for determining the large scale velocity 
structures appearing in vortex tangle, it is not sensitive for small scale Kelvin waves that 
have characteristic scale smaller than the intervortex separation. This is illustrated in 
Fig.~\ref{f.EkVortexKWs} where we have plotted the 3D energy spectrum for a straight vortex 
with and without Kelvin waves. The $1/k$ contribution coming from the straight vortex 
is so dominant that very little can be said about the Kelvin spectrum, at least for spectra 
with $\eta > 1.5$. The increased length only raises the spectrum up without affecting the slope 
in the high-$k$ region. In case of vortex ring the identification of Kelvin waves is even more 
complicated due to the oscillations that are already present for a ring without Kelvin waves.
This just emphasizes that in order to identify the Kelvin spectrum the direct measurement of 
the vortex position becomes more important than determination of the 3D energy spectrum at 
high $k$-values.

\section{Fractal dimension}

At low temperatures vortices typically look quite wiggly, almost fractal like. 
This is due to Kelvin waves. Therefore, one way to characterize the tangle is to determine the 
fractal dimension for vortex lines\cite{BaggaleyPRB2011}. 

Upper-bound estimate for the fractal dimension can be calculated by determining the correlation 
dimension $D$ introduced by Grassberger and Procaccia \cite{Grassberger1983}. In case of $N$ 
discretization points, $D$ can be determined by calculating the number of point pairs, $n$, 
whose separation is less than $\epsilon$. In the limit of $\epsilon\rightarrow 0$ the 
correlation integral $K(\epsilon) = n/N^2$ takes the form $K(\epsilon)\propto \epsilon^D$. 

\begin{figure}[!ht]
\begin{center}
\includegraphics[%
  width=0.9\linewidth,
  keepaspectratio]{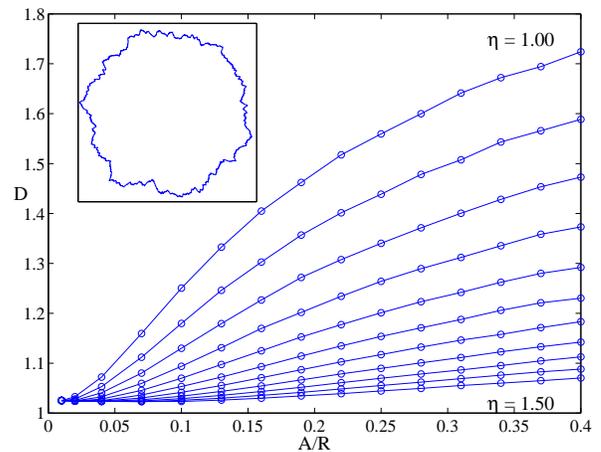}
\end{center}
\caption{(Color online) Correlation dimension $D$ as a function a Kelvin-wave amplitude for a
vortex ring occupied by Kelvin waves $|m| \ge 6$ for different spectra $|w_{\rm m}| \propto m^{-\eta}$ 
with $\eta = 1.00, 1.05, \ldots 1.50$ and using 4096 points. Calculated $D$ is an average over 
several samples where the phase and the distribution of positive and negative Kelvin modes is 
randomly set. The inset shows one configuration when $\eta = 1.2$ and $A/R = 0.25$. 
}
\label{f.fdim}
\end{figure}

How are different Kelvin spectra and $D$ related? In order to answer this  
we have determined $D$ in case of vortex ring that is occupied by Kelvin waves with different
spectra and with different amplitudes. Figure \ref{f.fdim} summarizes our results. As expected, 
$D$ is clearly bigger than unity only when $\eta \lesssim 1.5$ and when the amplitude is large 
enough. This limits the usability of the fractal dimension to spectra that are less steep than 
predicted theoretically. The dependence on the amplitude makes it difficult to determine $\eta$
accurately. However, if one observes a correlation dimension that is clearly larger than unity,
then one can almost certainly state that $\eta < 1.5$. Also, even if we used a simple vortex
ring to determine the fractal dimension, the results should be valid more generally at scales
smaller than the average intervortex separation, {\it i.e.} in the region where the Kelvin 
cascade is important.   

The difficulty in determining the fractal dimension is that numerical noise near the resolution 
limit (appearing because there are too few pairs) causes that $D \rightarrow 3$. We observed 
that this happens especially when the amplitude is large and $\eta$ is small (near unity). Therefore, 
the fit region should be limited to scales larger than the numerical resolution. The incorrectly 
chosen fit region might explain why the calculated correlation dimensions (of order 1.5) in 
Ref.~[\onlinecite{BaggaleyPRB2011}] are much bigger than the presented configuration (Fig.~20 in that
article) would suggest. For example, Kelvin spectrum with $\eta = 1.20$ and with amplitude 
$A = 0.25R$, presented in the inset of Fig.~\ref{f.fdim}, results in a fractal dimension of 
$D \approx 1.20$ only.

\section{Smoothed vortex}

\begin{figure}[!ht]
\begin{center}
\includegraphics[width=\linewidth,keepaspectratio]{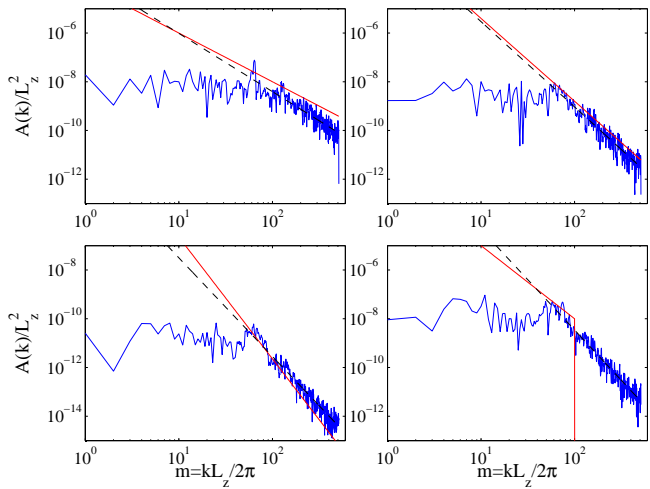}
\end{center}
\caption{(Color online) Kelvin-amplitude spectrum $A(k)$ calculated using a ``smoothed vortex''
 -method described in the text, with 1024 points and $n=16$. The (red) solid lines describe the 
correct spectrum and the dashed lines are fits at large $k$, where the method should be applicable. 
{\it Top left}: spectrum with $\beta=2$ (slope for the fit is 2.4).
{\it Top right}: spectrum with $\beta=3.4$ (fit 3.5).
{\it Bottom left}: spectrum with $\beta=5$ (fit 4.0).
{\it Bottom right}: spectrum with $\beta=3$, but omitting all modes with $m>100$ (fit gives 4.1).
} 
\label{f.smooth}
\end{figure}

For a vortex tangle the identification of Kelvin waves can be done by using the concept of  
``smoothed vortex''\cite{SvistunovPRB1995,BaggaleyPRB2011}. From the original discretized vortex
filaments, determined by points ${\bf s}_i$ ($i=1,\ldots,N$), one can construct a smoothed 
vortex line ${\bf s}_{\rm smooth}$ by using every $n$th points as nodes for a cubic-spline 
interpolation. One can then define the Kelvin-wave amplitude $a(\xi)$ as the distance 
between the smoothed and the original filament: $a(\xi) = |s-s_{\rm smooth}|$. The amplitude 
spectrum $A(k)$ can be defined as 
\begin{equation}
\frac{1}{2}\int a^2(\xi)d\xi = \int_0^\infty A(k)dk \, . 
\end{equation}
We have applied this method for a straight vortex with Kelvin waves, 
where the spectrum takes some known form. Figure~\ref{f.smooth} illustrates that
the above scheme has a tendency to follow the correct spectrum. For type 
$A(k)\propto k^{-\beta}$ spectrum, $\beta$ obtained from the fit is typically too
big for spectra with $\beta \lesssim 4$, and too small for spectra with $\beta \gtrsim 4$.
In some cases the method can result in a spectrum that is totally incorrect. This is also 
shown in Fig.~\ref{f.smooth}.
Better identification of Kelvin waves could perhaps be obtained if the smoothed vortex is allowed 
to avoid the original data points. This removes the discontinuities of the derivatives $a'(\xi)$ 
appearing when $a(\xi)=0$, which is one reason why the spectrum tends towards $\beta=4$. We have
tried this type of interpolation but so far the improvements are only minor.

\section{Conclusions}

Here we have presented several methods that can be used to identify Kelvin waves. In case of a straight 
vortex or a vortex ring the identification can be done most reliable. However, in case of vortex
tangle a great care should be used. The biggest problem is the lack of proper definition of a 
Kelvin wave on a curved vortex. 
In order to properly define a Kelvin wave there must exist a scale 
separation between the characteristic size of the underlying vortex and its disturbances.

Generally the Kelvin cascade to smaller scales can be qualitatively 
identified using, {\it e.g.} average curvature. An accurate determination of the Kelvin amplitude is
still very demanding and can result in large errors, depending on the numerical method used. 
It is alarming that some numerical methods can result in a spectrum that is totally 
incorrect but still very close to the spectrum predicted theoretically. Therefore, all methods 
should be properly tested using a simple configuration where the spectrum is known in advance.

The highly dominating geometrical $1/k$ scaling in the 3D energy spectrum at high-$k$ illustrates 
that at small scales the determination of the vortex location becomes much more essential than
the knowledge of the energy spectrum. Only after determining the vortex location, we have some
hope for identifying the Kelvin spectrum. Identifying the vortex location is not a problem with
the vortex filament model and is quite straight forward with the Gross-Pitaevskii equation.
Experimentally the identification is challenging but recently the visualization of quantized vortices has 
become possible\cite{LathropNature2006}.

Furthermore, the accuracy of the numerical schemes typically limits the proper determination of the 
Kelvin spectrum to scales $k \lesssim k_{\rm res}/5$, or even less. 
Similar requirement results from the dissipation, which must be present in order to avoid 
numerical bottleneck that would otherwise generate a non-physical fractalization of the vortex 
at the smallest scales.
Therefore, the identification of 
Kelvin waves on a tangle is a great challenge and requires resolution that is much higher than the 
average vortex separation. Here faster computers and especially new numerical algorithms, like the 
tree method\cite{BaggaleyJLTP2012tree}, are of great importance. \\
  

\begin{acknowledgments}
We acknowledge the support from the Academy of Finland and EU 7th Framework Programme (FP7/2007-2013, 
Grant 228464, MicroKelvin).
\end{acknowledgments}


\end{document}